\documentclass[pra,twocolumn,amsmath,amssymb,superscriptaddress]{revtex4}

\usepackage{graphicx}
\usepackage{dcolumn}
\usepackage{bm}
\usepackage{color}

\usepackage{epsfig,float,afterpage,amssymb,wrapfig,psfrag}
\usepackage{subfigure}

\newcommand{\be}{\begin{equation}}
\newcommand{\bea}{\begin{eqnarray}}
\newcommand{\bk}{\mathbf{k}}
\newcommand{\bq}{\mathbf{q}}
\newcommand{\br}{\mathbf{r}}
\newcommand{\ee}{\end{equation}}
\newcommand{\eea}{\end{eqnarray}}
\newcommand{\half}{\text{$\textstyle\frac{1}{2}$}}
\newcommand{\pdag}{\phantom{\dag}}

\begin{document}
\title{Pairing correlations near a Kondo-destruction quantum critical point}
\author{J.\ H.\ Pixley}
\affiliation{Condensed Matter Theory Center and Joint Quantum Institute,
Department of Physics, University of Maryland, College Park,
Maryland 20742-4111, USA}
\affiliation{Department of Physics and Astronomy, Rice University,
Houston, Texas 77005-1827, USA}
\author{Lili Deng}
\affiliation{Department of Physics, University of Florida,
Gainesville, Florida 32611-8440, USA}
\author{Kevin Ingersent}
\affiliation{Department of Physics, University of Florida,
Gainesville, Florida 32611-8440, USA}
\author{Qimiao Si}
\affiliation{Department of Physics and Astronomy, Rice University,
Houston, Texas 77005-1827, USA}

\date{\today}
\begin{abstract}
Motivated by the unconventional superconductivity observed in heavy-fermion
metals, we investigate pairing susceptibilities near a continuous quantum phase
transition of the Kondo-destruction type. We solve two-impurity Bose-Fermi
Anderson models with Ising and Heisenberg forms of the interimpurity exchange
interaction using continuous-time quantum Monte-Carlo and numerical
renormalization-group methods. Each model exhibits a Kondo-destruction quantum
critical point separating Kondo-screened and local-moment phases. For
antiferromagnetic interimpurity exchange interactions, singlet pairing is
found to be enhanced in the vicinity of the transition. Implications of this
result for heavy-fermion superconductivity are discussed.
\end{abstract}


\maketitle

A quantum critical point (QCP) arises when matter continuously transforms
from one ground state to another \cite{JLTP-issue10}.
Whether and how a magnetic QCP underlies unconventional superconductivity in
correlated electron systems  remains one of the central questions in condensed
matter physics \cite{Coleman05,QS.10,Mathur98}.
At a macroscopic level, a QCP is accompanied by an enhanced entropy \cite{Wu10}.
At sufficiently low temperatures, in the proximity of a QCP, it is natural for
the enhanced entropy to promote emergent phases such as superconductivity. At
a microscopic level, however, how quantum criticality drives superconductivity
remains an open issue. Developing an understanding of unconventional
superconductivity is pertinent to a large list of correlated materials such as
iron pnictides, copper oxides, organics, and heavy fermions.

An important opportunity for detailed exploration of this general
issue is provided by heavy-fermion metals, in which many QCPs
have been explicitly identified \cite{QS.10,HVL.07}. Theoretical studies
have shown that antiferromagnetic QCPs in a Kondo lattice system fall into two
classes. Spin-density-wave QCPs are described in the Landau framework
of order-parameter fluctuations \cite{Hertz.76+Millis.93}. The other class of
QCPs goes beyond the Landau approach by invoking a critical destruction of the
Kondo effect \cite{Si.01+Si.03,Coleman.01}. Distinctive features of this ``local
quantum criticality'' include $\omega/T$ scaling in the spin susceptibility and
the single-particle spectral function, vanishing of an additional energy scale,
and a jump in the Fermi-surface volume. There is mounting experimental evidence
for these characteristic properties, e.g., from inelastic neutron-scattering
measurements on Au-doped CeCu$_6$ \cite{Aronson.95+Schroeder.00}, scanning
tunneling spectroscopy on CeCoIn$_5$ \cite{Aynajian.12}, Hall-effect and
thermodynamic measurements on YbRh$_2$Si$_2$
\cite{Paschen.00+Gegenwart.07+Friedemann.10}, and magnetic quantum-oscillation
measurements on CeRhIn$_5$ \cite{Shishido}.

Given the considerable advances in the understanding of the unconventional
quantum critical behavior of heavy fermions in the normal state, it is clearly
important to address its implications for superconductivity. Theoretically, it
remains an open question whether a Kondo-destruction QCP promotes unconventional
superconductivity \cite{Gegenwart.08}. To make progress, it is essential to
identify simplified models in which this issue can be addressed and insights
can be gained. Because an on-site Coulomb repulsion does not favor conventional
$s$-wave pairing, this issue can only be studied in models that incorporate
correlations among different local-moment sites.

In this work, we propose perhaps the simplest models that support
Kondo-destruction physics and allow the study of superconducting correlations:
two local moments that interact with each other through a direct exchange
interaction and are also coupled both to a conduction-electron band and to a
bosonic bath. The models we have considered can be obtained from a cluster
generalization of the extended dynamical mean field theory (C-EDMFT)
\cite{Pixley.14} applied to the periodic Anderson model with Ising anisotropy.
The critical physics arises from the antiferromagnetic channel, which we will
be concerned with. We then arrive at the model defined in Eq.\ \eqref{Ham}
below \cite{Pixley.14}. In the past, significant insights have been gained from
single-impurity models, where Kondo-destruction QCPs are characterized by a
vanishing Kondo energy scale, an $\omega/T$ scaling in
the local spin susceptibility, and a linear-in-temperature single-particle
relaxation rate \cite{Glossop.05,Ingersent,Kirchner,Pixley.10,pixley,Pixley.13}.
Such properties are reminiscent of the aforementioned experiments near the
antiferromagnetic QCPs of heavy-fermion metals.

We solve the two-impurity Bose-Fermi Anderson models via a continuous-time
quantum Monte Carlo (CT-QMC) approach \cite{Pixley.10,Pixley.13,Gull} and
using the numerical renormalization group (NRG) \cite{Wilson.75,Glossop.05}.
We determine the magnetic quantum critical properties and compute pairing
susceptibilities across the phase diagram. We find that pairing correlations
are in general enhanced near the Kondo-destruction QCP. This suggests a new
mechanism for superconductivity near antiferromagnetic quantum phase
transitions (QPTs).

\begin{figure}[t!]
\includegraphics[height=1.25in]{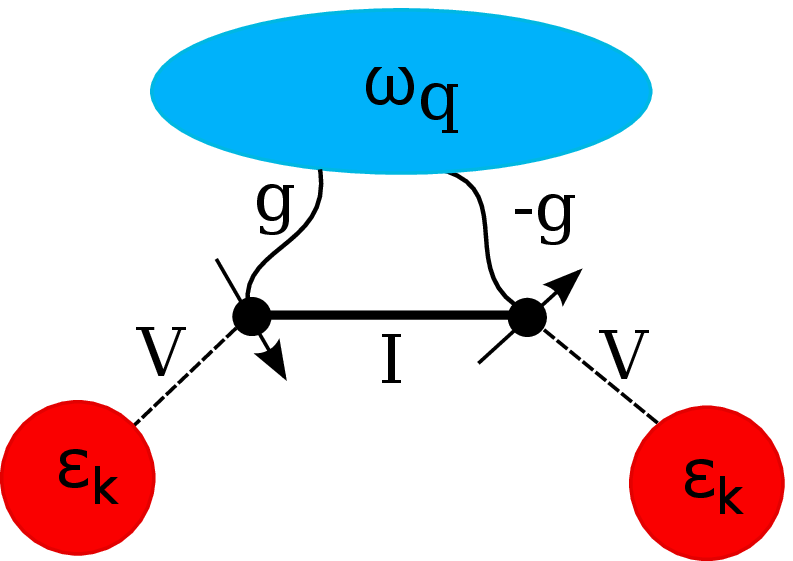}
\caption{\label{fig:model}(Color online)
Schematic representation of the two-impurity Bose-Fermi Anderson models
considered in this work. The impurity spins interact via a direct exchange
coupling $I$ (or $I_z$), and $S_1^z-S_2^z$ couples with strength $g$ to a
dissipative bosonic bath having dispersion $\omega_{\bq}$.
For very large impurity separation, each impurity effectively hybridizes with
strength $V$ with its own conduction band of dispersion $\epsilon_{\bk}$.}
\end{figure}

The two-impurity Bose-Fermi Anderson models, illustrated in Fig.\
\ref{fig:model}, are defined by Hamiltonians of the form
\begin{align}
H =
&\sum_{i=1,2} \biggl( \epsilon_d \sum_{\sigma} n_{di\sigma}
  +  U n_{d i \uparrow}n_{d i \downarrow} \biggr)  + H_{12} \notag \\
&+ \sum_{{\bk},\sigma}\epsilon_{\bk}c_{\bk\sigma}^{\dag} c_{\bk\sigma}^{\pdag}
  + \frac{V}{\sqrt{N_k}} \sum_{i,\bk,\sigma} \left(e^{i\bk\cdot\br_i}
  d_{i\sigma}^{\dag} c_{\bk\sigma}^{\pdag} +\mathrm{H.c.} \right) \notag \\
&+ \sum_{\bq} \omega_{\bq} \phi_{\bq}^{\dag} \phi_{\bq}^{\pdag}
  + g(S_1^z - S_2^z) \sum_{\bq}(\phi_{\bq}^{\dag} + \phi_{-\bq}^{\pdag}) .
\label{Ham}
\end{align}
Here, $d_{i\sigma}$ destroys an electron on impurity site $i=1$ or $2$ with
spin $\sigma = \,\uparrow$ or $\downarrow$, energy $\epsilon_d$, and on-site
Coulomb repulsion $U$; $n_{di\sigma}=d_{i\sigma}^{\dag} d_{i\sigma}^{\pdag}$,
and $\mathbf{S}_i=\half\sum_{\alpha,\beta}d_{i\alpha}^{\dag}
\bm{\sigma}_{\alpha\beta}d_{i\beta}^{\pdag}$ where
$\sigma_{\alpha\beta}^{x,y,z}$ are the Pauli matrices. The operator
$c_{\bk\sigma}$ destroys a conduction electron with wave vector $\bk$, spin
$\sigma$, and energy $\epsilon_{\bk}$ that has a hybridization $V$ with each
impurity, while $\phi_{\bq}$ destroys a boson with energy $\omega_{\bq}$ that
couples with strength $g$ to the difference of impurity spin $z$ components.
$N_k$ is the number of $\bk$ values.

To control the interimpurity exchange interaction, we take the limit of infinite
impurity separation $|\br_1 - \br_2|$ to ensure the vanishing of the indirect
Ruderman-Kittel-Kasuya-Yosida exchange interaction between $\mathbf{S}_1$ and
$\mathbf{S}_2$. Then impurities $1$ and $2$ hybridize with linearly independent
combinations of band states, and interact only through their coupling to the
bosonic bath and via a direct exchange term $H_{12}$, either of the Ising form
$H_{12}=I_z S_1^zS_2^z$ or the Heisenberg form
$H_{12}=I\mathbf{S}_1\cdot\mathbf{S}_2$. Ising exchange is naturally obtained
from the C-EDMFT approach~\cite{Pixley.14}, but including the static Heisenberg
interaction allows us to study anisotropic couplings since this breaks the purely
Ising coupling of the bosonic bath. We note that integrating out the bosonic
bath will induce a retarded antiferromagnetic exchange of Ising symmetry.

We assume a flat electronic density of states
$\rho_c(\epsilon) = \rho_0\Theta(D-|\epsilon|)$
and a sub-Ohmic bosonic density of states
\be
\rho_{\phi}(\omega) =
  K_0^2\omega_c^{1-s}\omega^s \, \Theta(\omega) \, f(\omega/\omega_c).
\label{dos}
\ee
For the CT-QMC calculations we have used a cutoff function $f(x)=\exp(-|x|)$
and chosen $K_0^{-2} = \omega_c^{2}\Gamma(s+1)$ so that the density of states
is normalized to unity. Within the NRG, we use $f(x)=\Theta(1-|x|)$ with
$K_0=1$. In this work we restrict ourselves to the range $1/2<s<1$.

In the absence of the bosonic bath, the pure-fermionic two-impurity Anderson
model can be mapped via a Schrieffer-Wolff transformation to a two-impurity
Kondo model with a direct exchange interaction \cite{tzen}. In the case of
Heisenberg exchange, both the Anderson and Kondo formulations are well studied
\cite{Jones,Affleck}, displaying a critical point at an antiferromagnetic
exchange $I_c>0$ in the presence of particle-hole symmetry; at this point,
the static singlet-pairing susceptibility diverges \cite{Zhu}. For an Ising
$H_{12}$, the model possesses a Kosterlitz-Thouless (KT) QPT at $|I_z^c|>0$
between a Kondo-screened phase and an interimpurity Ising-ordered phase
\cite{Andrei,Garst}. Without the conduction band, Eq.\ \eqref{Ham}
reduces to a two-spin boson model; studies of this model with $S_1^z+S_2^z$
coupled to a spin bath found a QCP separating a delocalized phase and
a ferromagnetically localized phase \cite{Orth,McCutcheon}.

We have solved Eq.\ \eqref{Ham} with $H_{12}$ of Ising form by extending the
CT-QMC approach \cite{Gull,Pixley.10,Pixley.13}. At temperature
$T = 1/\beta > 0$ we determine the staggered Binder cumulant
\cite{Binder.81,pixley,Pixley.13}
$U_4^s(\beta,I_z,g)=\langle M_s^4 \rangle/\langle M_s^2\rangle^2$,
where the staggered magnetization
$M_s = \beta^{-1} \int_0^{\beta} d\tau\,S_s^z(\tau)$ with
$S_s^z=\half(S_1^z-S_2^z)$, and the staggered static spin susceptibility
$\chi_{s}(T) = \beta\langle M_s^2 \rangle$.
To solve the Heisenberg form of $H_{12}$ we have used the Bose-Fermi extension
\cite{Glossop.05} of the NRG \cite{Wilson.75,Bulla.08}.
To measure the pairing correlation between the $d$ electrons at different
impurity sites, we study dynamic singlet ($d$-wave) and triplet
($p$-wave) pairing susceptibilities
\be
\chi_{\alpha}(\tau,\beta)
  = \langle T_{\tau} \Delta_{\alpha}(\tau) \Delta_{\alpha}^{\dag}\rangle,
  \;\; \alpha=d \text{ or } p,
\label{pairing}
\ee
where
$\Delta_d^{\dag}
  \!=\!\frac{1}{\sqrt{2}}(d_{1\uparrow}^{\dag}d_{2\downarrow}^{\dag}
  \!- d_{1\downarrow}^{\dag}d_{2\uparrow}^{\dag})$,
$\Delta_p^{\dag}
  \!=\! \frac{1}{\sqrt{2}}(d_{1\uparrow}^{\dag}d_{2\uparrow}^{\dag}
  \!+ d_{1\downarrow}^{\dag}d_{2\downarrow}^{\dag})$,
and $T_{\tau}$ orders in imaginary time.
The static pairing susceptibilities follow via
$\chi_{\alpha}(T)=\int_0^{\beta}d\tau\,\chi_{\alpha}(\tau,\beta)$.
Each numerical technique as applied to the models studied here
is further described in the Supplemental Material \cite{suppl}, and
additional details will be given elsewhere \cite{long_paper}.

\begin{figure}[t!]
\begin{minipage}[b]{10pc}
\includegraphics[height=1.75in,angle=-90]{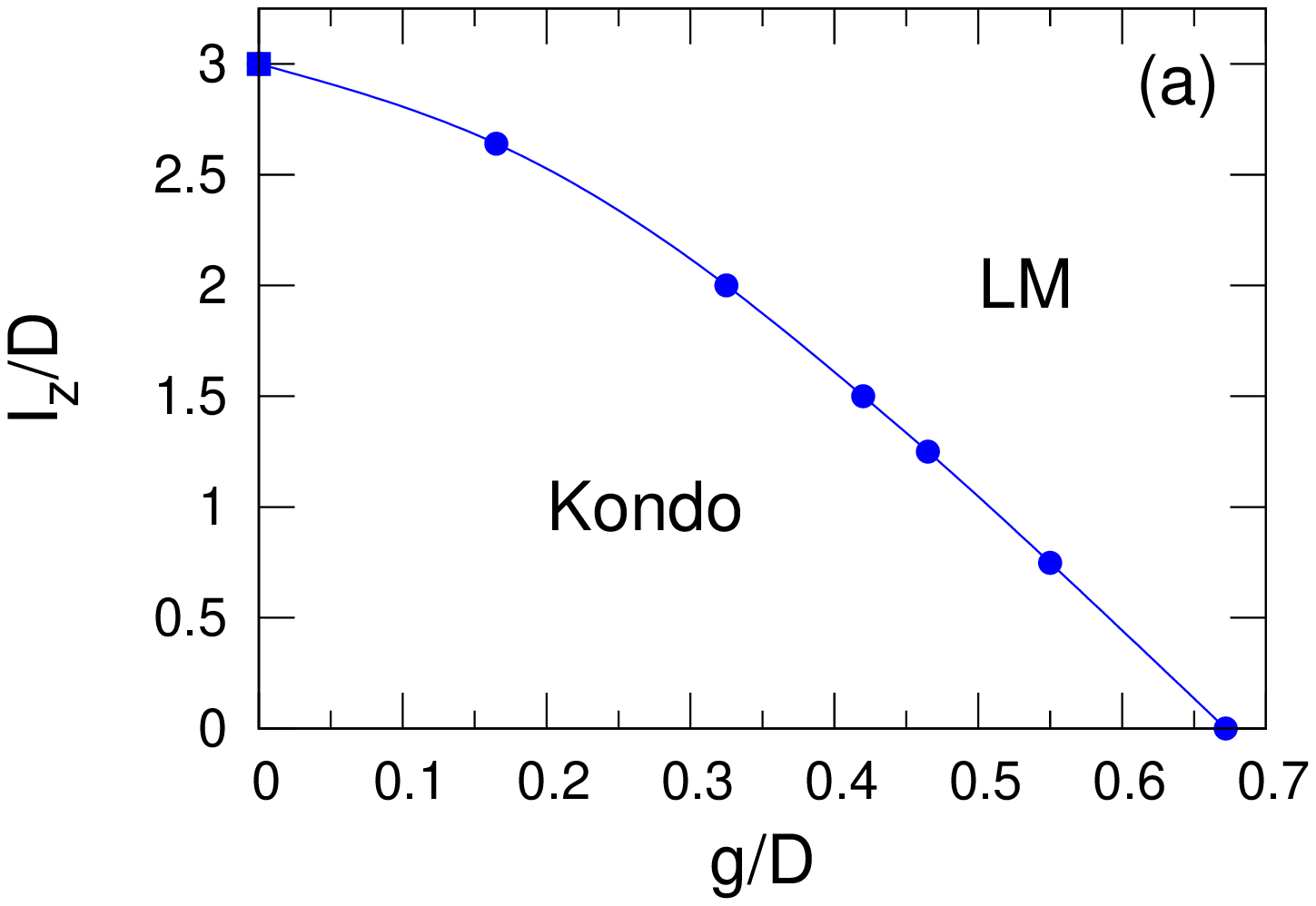}
\end{minipage}
\begin{minipage}[b]{10pc}
\includegraphics[height=1.75in,angle=-90]{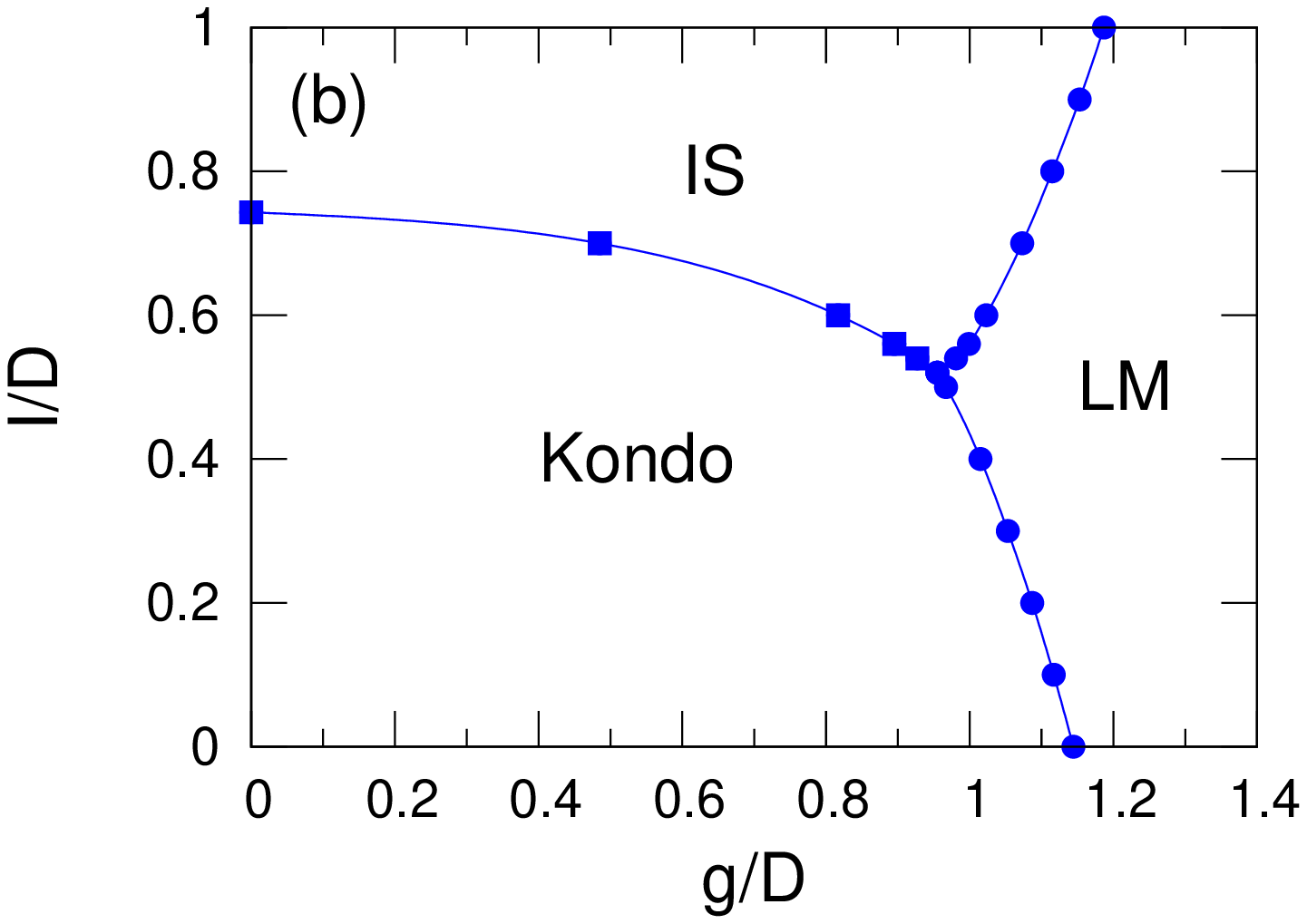}
\end{minipage}
\caption{\label{fig:phase_diagram}(Color online)
(a) Phase diagram for an Ising $H_{12}$ determined via CT-QMC from the
staggered Binder cumulant. A square marks the Kosterlitz-Thouless QPT and
circles denote second-order Kondo-destruction QPTs governed by the QCP at
$I_z=0$.
(b) Phase diagram for a Heisenberg $H_{12}$ found with the NRG. Squares
represent QPTs governed by the critical point at $g=0$, while circles
represent Kondo-destruction QPTs induced by the coupling to the bosonic
bath. Kondo-screened (Kondo), interimpurity-singlet (IS), and local-moment
(LM) phases all meet at a tricritical point.}
\end{figure}

In the following, we work with fixed $\Gamma_0=0.25D$ and
$U=-2\epsilon_d=0.001D$. This choice places the Anderson impurities at mixed
valence with a high Kondo temperature $T_K\simeq 1.39 D$ (for $g,I_z,I=0$),
ensuring a correspondingly high temperature of entry into the quantum critical
regime \cite{Pixley.13}. We also take $\omega_c=D$ and focus on sub-Ohmic bath
exponents [see Eq.\ \eqref{dos}] $s=0.8$ for Ising exchange and $s=0.6$
for the Heisenberg case.

\begin{figure}[t!]
\begin{minipage}[b]{10pc}
\includegraphics[height=1.75in,angle=-90]{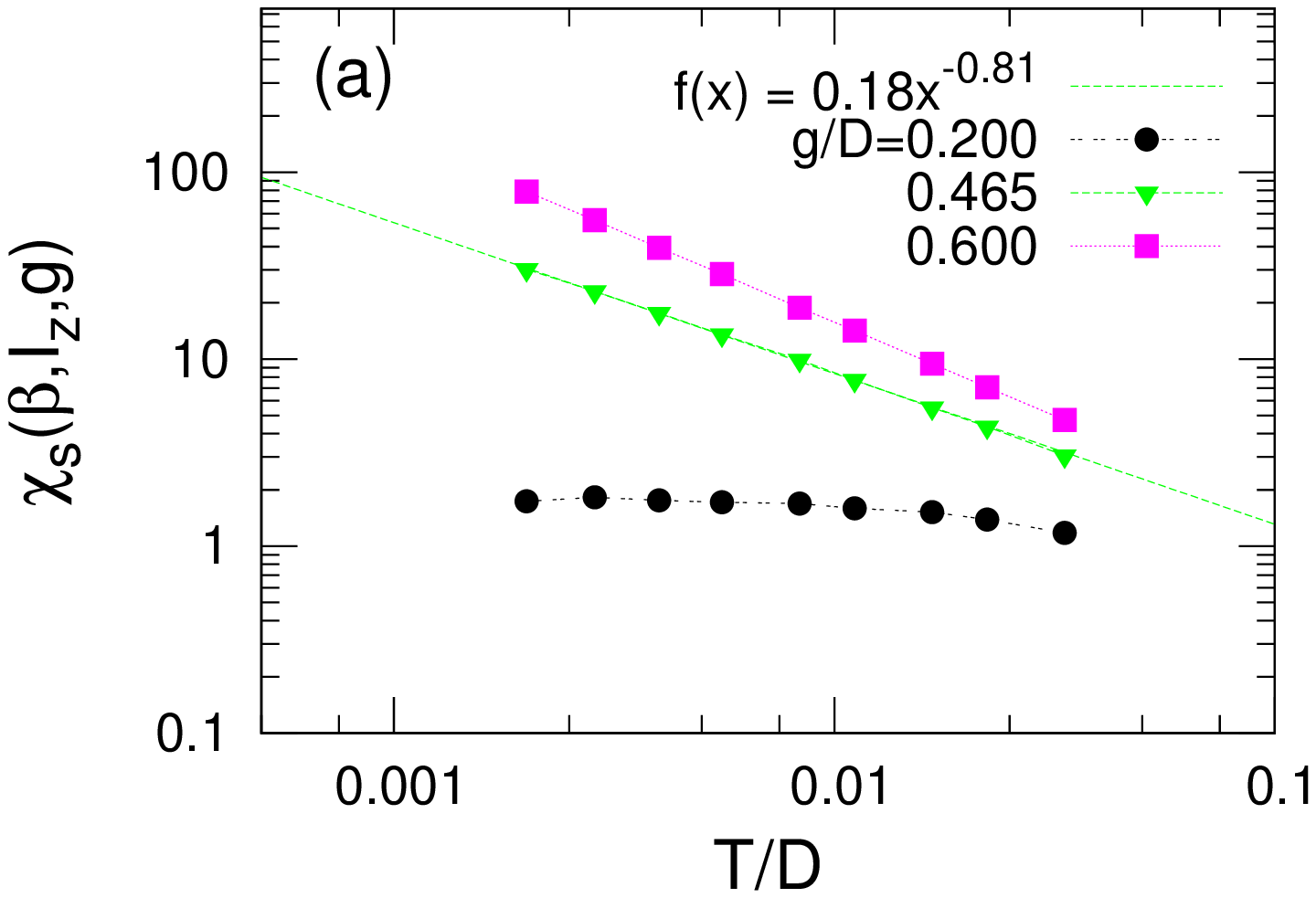}
\end{minipage}
\begin{minipage}[b]{10pc}
\includegraphics[height=1.75in,angle=-90]{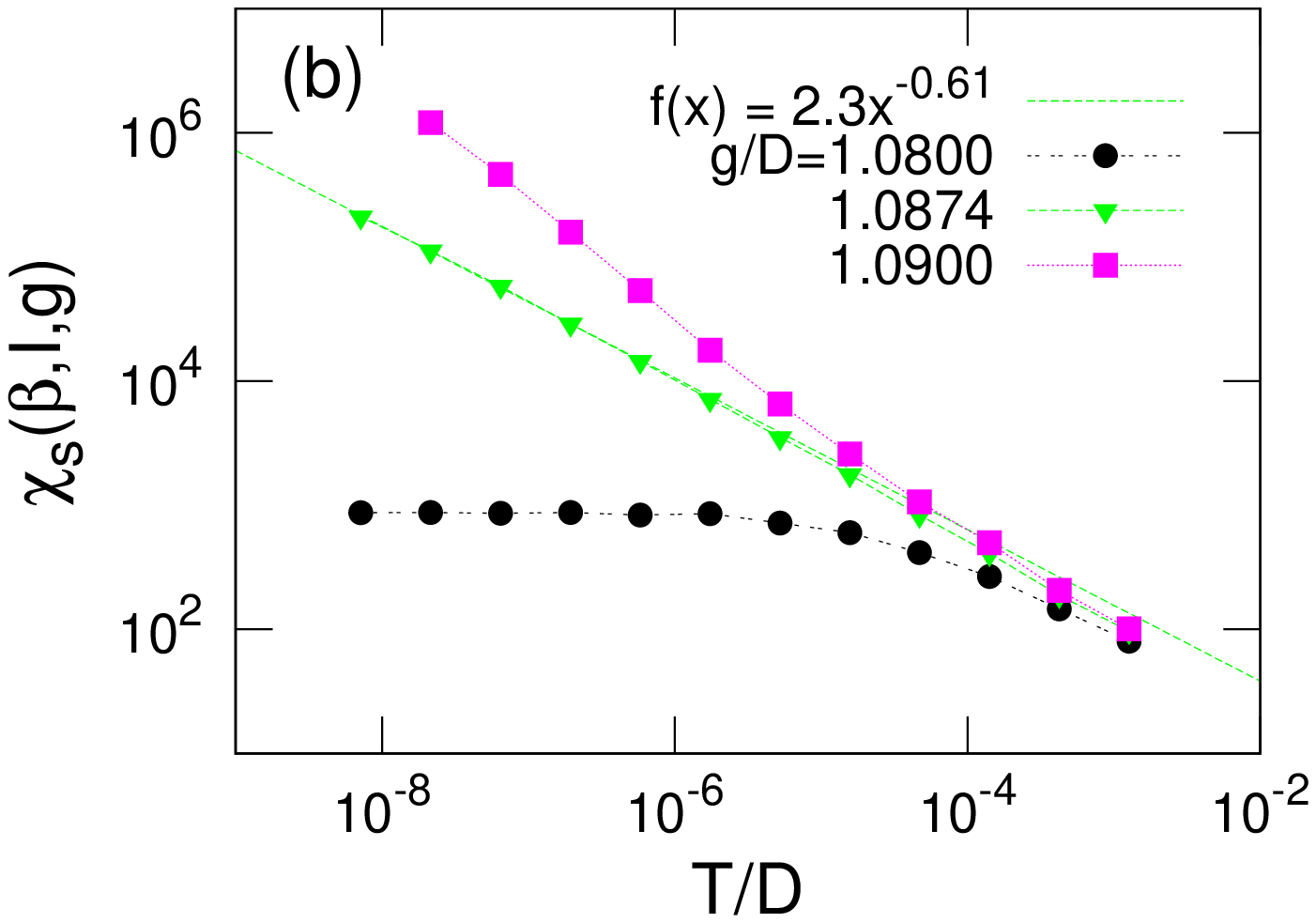}
\end{minipage}
\caption{\label{fig:chi_stat}(Color online)
Staggered spin susceptibility $\chi_s(T)$ in the Kondo phase (circles),
at the QCP (triangles), and in the LM phase (squares), for
(a) an Ising $H_{12}$ with $s=0.8$,
and (b) a Heisenberg $H_{12}$ with $s=0.6$.
At the QCP, $\chi_s \sim T^{-s}$.}
\end{figure}

\begin{figure}[t!]
\begin{minipage}[b]{10pc}
\includegraphics[height=1.7in, angle=-90]{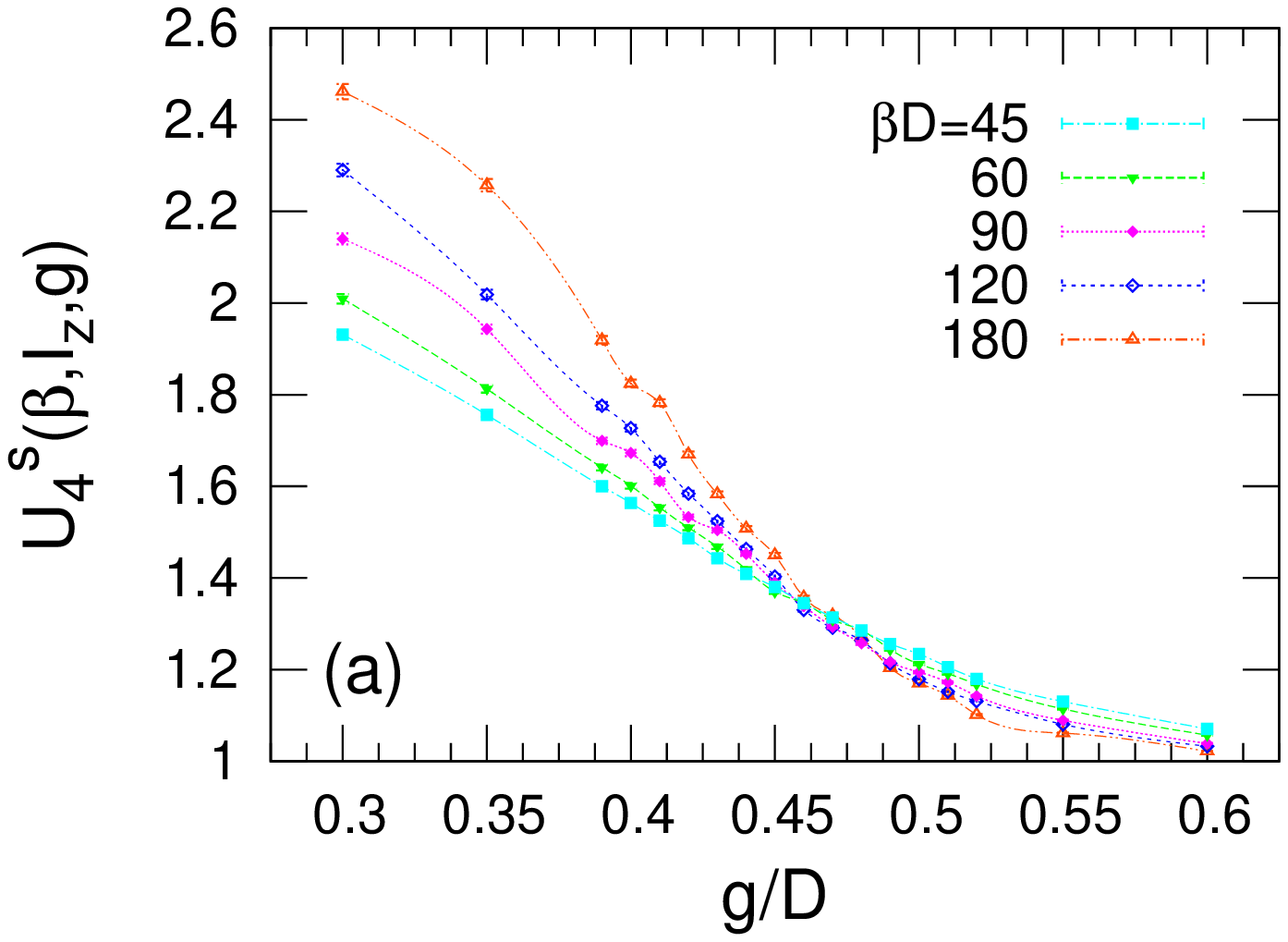}
\end{minipage}
\begin{minipage}[b]{10pc}
\includegraphics[height=1.8in, angle=-90]{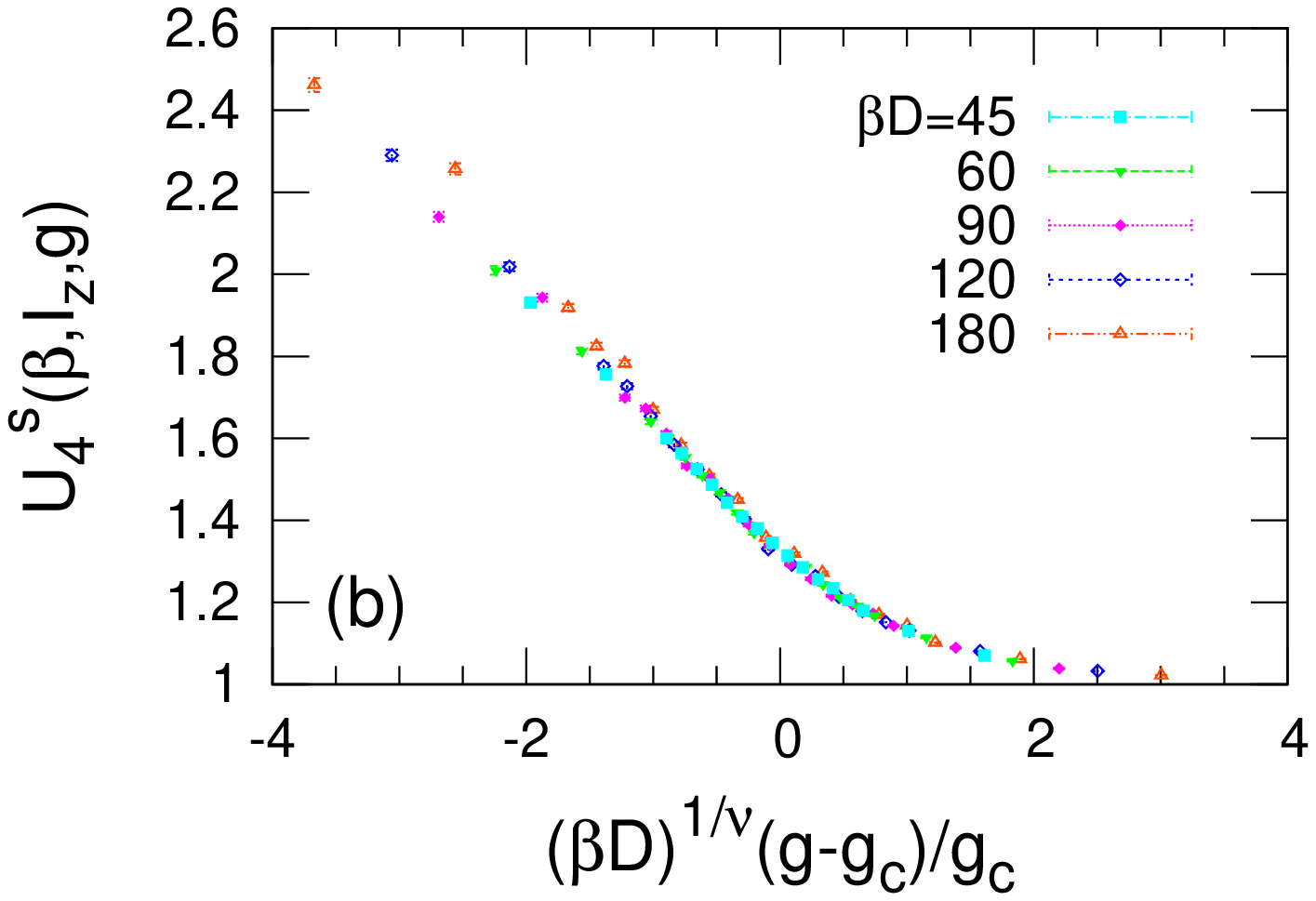}
\end{minipage}
\newline
\begin{minipage}[b]{10pc}
\includegraphics[height=1.7in, angle=-90]{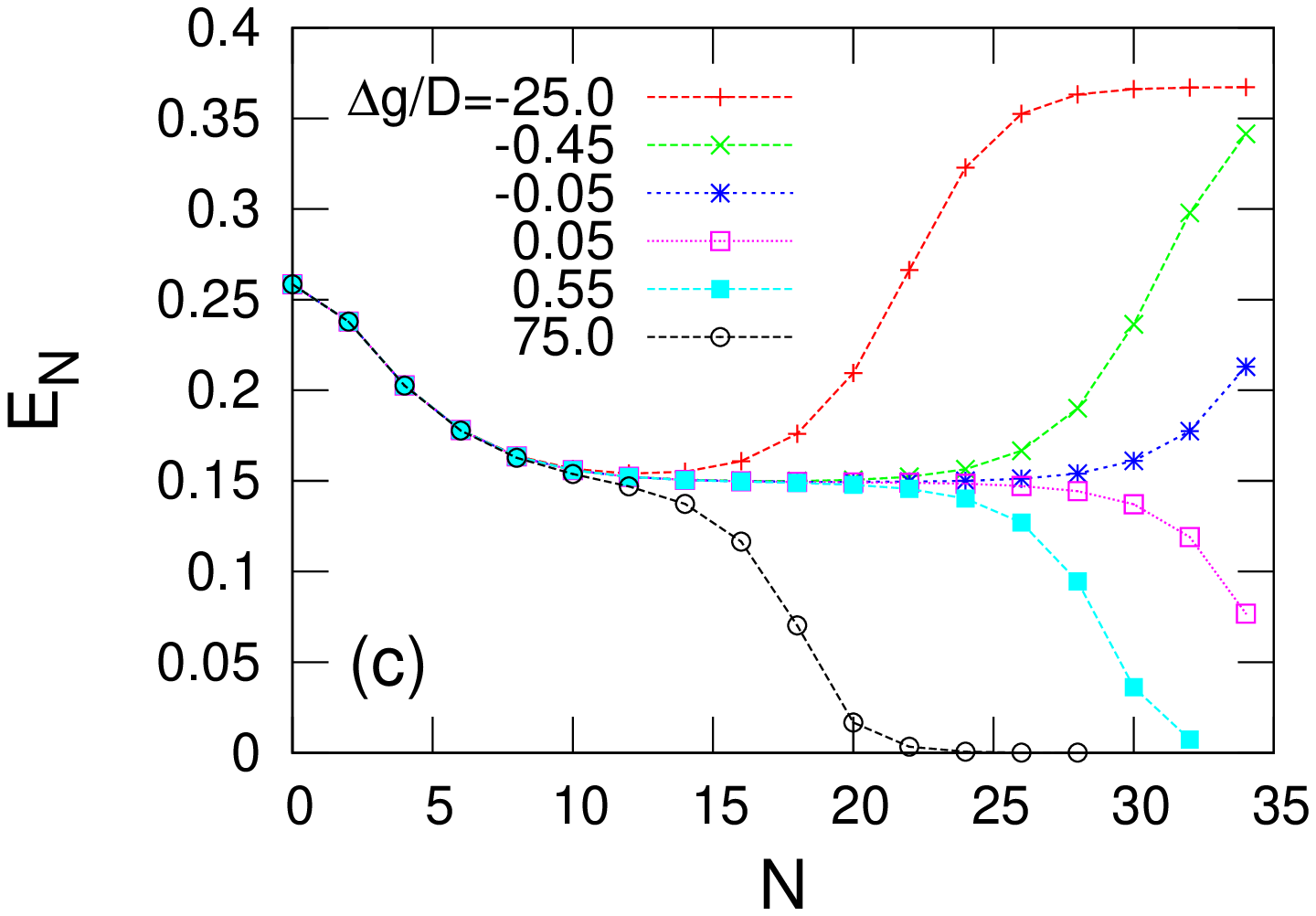}
\end{minipage}
\begin{minipage}[b]{10pc}
\includegraphics[height=1.8in, angle=-90]{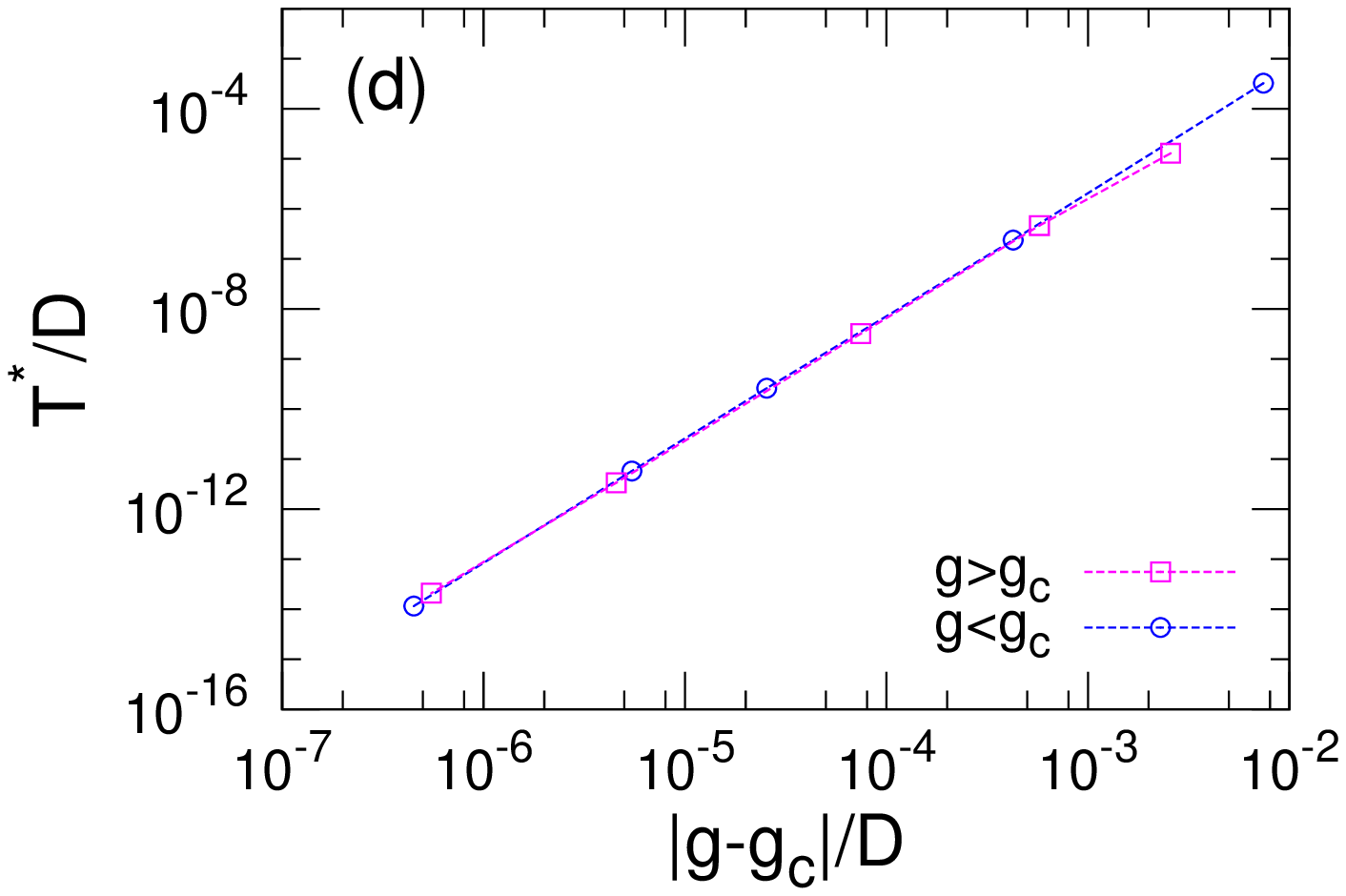}
\end{minipage}
\caption{\label{fig:near_gc}(Color online)
(a) Binder cumulant $U_4^s(\beta,I_z,g)$ vs $g$ for an Ising $H_{12}$ with
$I_z=1.25D$, $s=0.8$, and at the labeled inverse temperatures $1/\beta$.
The intersection of curves gives the
critical bosonic coupling $g_c/D=0.465(5)$.
(b) A scaling collapse of the same data near $g_c$ according to Eq.\
\eqref{bind_scal} yields a correlation-length exponent
$\nu(s=0.8)^{-1}=0.45(8)$.
(c) Flow of a low-energy NRG eigenstate vs iteration number $N$ for
a Heisenberg $H_{12}$ with $I=0.2D$, $s=0.6$, and six values of
$\Delta g \equiv 10^6(g-g_c)$, where $K_0 g_{c}=1.087\,425\,45(1)$.
(d) Low-energy crossover scale from the NRG, fitted to
$T^*\propto |g-g_c|^{\nu}$ yielding $\nu(s=0.6)^{-1}=0.40(2)$.}
\end{figure}

\emph{Ising $H_{12}$}. Figure \ref{fig:phase_diagram}(a) shows the $T=0$
phase diagram for the case of Ising exchange, as obtained using CT-QMC.
For $0\le g,\, I_z \ll D$, each impurity spin is locked into a Kondo singlet
with the conduction band and $\chi_s(T)$ approaches a constant at low
temperatures [e.g., Fig.\ \ref{fig:chi_stat}(a)].
Upon increasing $g$ and/or $I_z$, the system passes through a QPT into an
Ising-antiferromagnetic local-moment (LM) phase in which the impurity spins
are anti-aligned and decoupled from the conduction band, as seen through a
Curie-Weiss behavior of the staggered spin susceptibility:
$\chi_s(T)\sim T^{-1}$ [Fig.\ \ref{fig:chi_stat}(a)]. The Kondo energy scale
vanishes continuously on the Kondo side of the QPT, characteristic of a
Kondo-destruction QCP. The staggered Binder cumulant $U_4^s(\beta,I_z,g)$
varies from 3 deep in the Kondo phase to 1 far into the LM phase. For fixed
$I_z$, the cumulant near the QCP has a scaling form
\begin{equation}
U_4^s(\beta,I_z,g) = U_4^s\bigl[\beta^{1/\nu}(g/g_c -1);I_z\bigr],
\label{bind_scal}
\end{equation}
identifying $g_c$ as the point of temperature independence of $U_4^s$ vs $g$
[Fig.\ \ref{fig:near_gc}(a)]. Optimizing the scaling collapse according to
Eq.\ \eqref{bind_scal} gives a correlation-length exponent
$\nu(s=0.8)^{-1}=0.45(8)$ [Fig.\ \ref{fig:near_gc}(b)], close to the
value $0.469(1)$ found using the NRG for the single-impurity Ising-symmetry
Bose-Fermi Kondo model \cite{Glossop.05}.

For $g=0$, the Ising critical point is KT-like, characterized by a divergence
$\chi_s(T,I_z=I_z^c,g=0)\sim T^{-1}$. Consequently, the coupling $g$ has a
scaling dimension $[g] = (1-s)/2$ and is relevant for $s<1$.
This dictates a flow away from the KT fixed point along the phase boundary in
Fig.\ \ref{fig:phase_diagram}(a) toward the $I_z=0$ critical point \cite{suppl}.
Tuning $g$ to the boundary at fixed $I_z>0$, we find that the staggered spin
susceptibility diverges as
\begin{equation}
\chi_s\big[T,I_z,g=g_c(I_z)\big] \sim T^{-x}
\label{stat}
\end{equation}
with $x=0.79(3),0.78(3),0.80(3),0.82(3),0.82(3),0.83(4)$ for increasing $I_z$.
These values are consistent with
$x=s$,
suggesting that the staggered channel
exhibits the same critical properties as the single-impurity Ising-symmetric
Bose-Fermi Kondo model \cite{Glossop.05}.

\emph{Heisenberg $H_{12}$}.
For Heisenberg exchange, the NRG gives the phase diagram shown in Fig.\
\ref{fig:phase_diagram}(b), based on runs performed with a basis of up to
$N_b = 4$ bosons per site of the bosonic Wilson chain and retaining up to
$N_s = 800$ many-body eigenstates at the end of each iteration. For small $g$
and $I$, the model is in the Kondo phase. Tuning $I$ for $g=0$, we pass through
a critical point into an interimpurity singlet (IS) phase, in which the
impurity spins are locked into a singlet and decoupled from the conduction band.
At the particle-hole-symmetric critical point \cite{Jones,Affleck}, the
staggered spin susceptibility diverges as $\chi_s(T,I=I_c,g=0) \sim \ln(T_K/T)$.
Using the corresponding scaling dimension of the staggered impurity spin,
along with the scaling dimension of $\phi_{\bq}$, we determine that
the bosonic coupling has scaling dimension $[g]=-s/2$ and is
irrelevant for $s>0$. Indeed, we find that the NRG spectrum along the
phase boundary is independent of $g$ for small values of $g$ \cite{suppl},
indicating that the critical behavior is governed by the $g=0$ QCP.

For small $I>0$, tuning the bosonic coupling $g$ yields a QPT from the Kondo
phase to the LM phase [Fig.\ \ref{fig:near_gc}(c)].
The Kondo energy scale vanishes continuously on approach from the small-$g$
side of this Kondo-destruction QCP. At the QCP, the staggered spin
susceptibility obeys Eq.\ \eqref{stat} with $I_z$ replaced by $I$ and
$x=0.61(2)$ [Fig.\ \ref{fig:chi_stat}(b)], again consistent with $x=s$.
Nearby, a low-energy crossover temperature $T^*$ (equal to the effective Kondo
temperature for $g<g_c$) varies as $T^*\propto |g-g_c|^{\nu}$, yielding
for the data shown in Fig.\ \ref{fig:near_gc}(d) a correlation-length
exponent $\nu(s=0.6)^{-1}=0.40(2)$. However, we find that (unlike the global
phase diagram and the value of the exponent $x$), the value of $\nu$ is
sensitive to the NRG truncation of states. Increasing $N_b$ from 4 to 6 and $N_s$
from $800$ to $1\,200$ leads to a refinement of our estimate to
$\nu(s=0.6)^{-1}=0.51(4)$, within numerical error identical to the value
$\nu(s=0.6)^{-1}=0.509(1)$ found for the single-impurity Ising-symmetry
Bose-Fermi Kondo model \cite{Glossop.05}. We therefore conclude that the
Kondo-destruction QCPs for Ising and Heisenberg exchange fall within the same
universality class. In both Ising and Heisenberg cases, the Kondo-destruction
QCPs are insensitive to breaking of particle-hole symmetry via setting
$U\neq -2\epsilon_d$, as well as to a finite impurity separation
\cite{long_paper}.

We turn next to the transition between the IS and LM phases. Fixing $I$ at a
large value and tuning $g$, the bosonic bath decoheres and destroys the
interimpurity singlet state at a QCP, where we find similar critical properties
to those on the Kondo-LM boundary: $\chi_s$ diverging according to Eq.\
\eqref{stat} with $x=0.61(3)$, and for $N_b = 4$,
a correlation length exponent $\nu(s=0.6)^{-1}=0.40(2)$ indistinguishable from
the corresponding value found on the Kondo-LM boundary.

In the particle-hole symmetric case that is the focus of this Rapid
Communication, the Kondo, IS, and LM phases all meet at a tricritical point, as
shown in Fig.\ \ref{fig:phase_diagram}(b). Generic particle-hole asymmetry is
known to turn the Kondo-to-IS transition in Fig.\ \ref{fig:phase_diagram}(b)
into a crossover \cite{Jones,Affleck}, leaving only a single line of
Kondo-destruction QPTs.

\emph{Pairing susceptibilities}. We now consider the singlet and triplet
pairing susceptibilities defined in Eq.\ \eqref{pairing}. For both the Ising
and Heisenberg forms of the interimpurity exchange, the static triplet pairing
susceptibility $\chi_p$ (not shown) is reduced by any nonzero value of $g$,
$I_z$, or $I$.

\begin{figure}[t!]
\begin{minipage}[b]{10pc}
\includegraphics[height=1.15in]{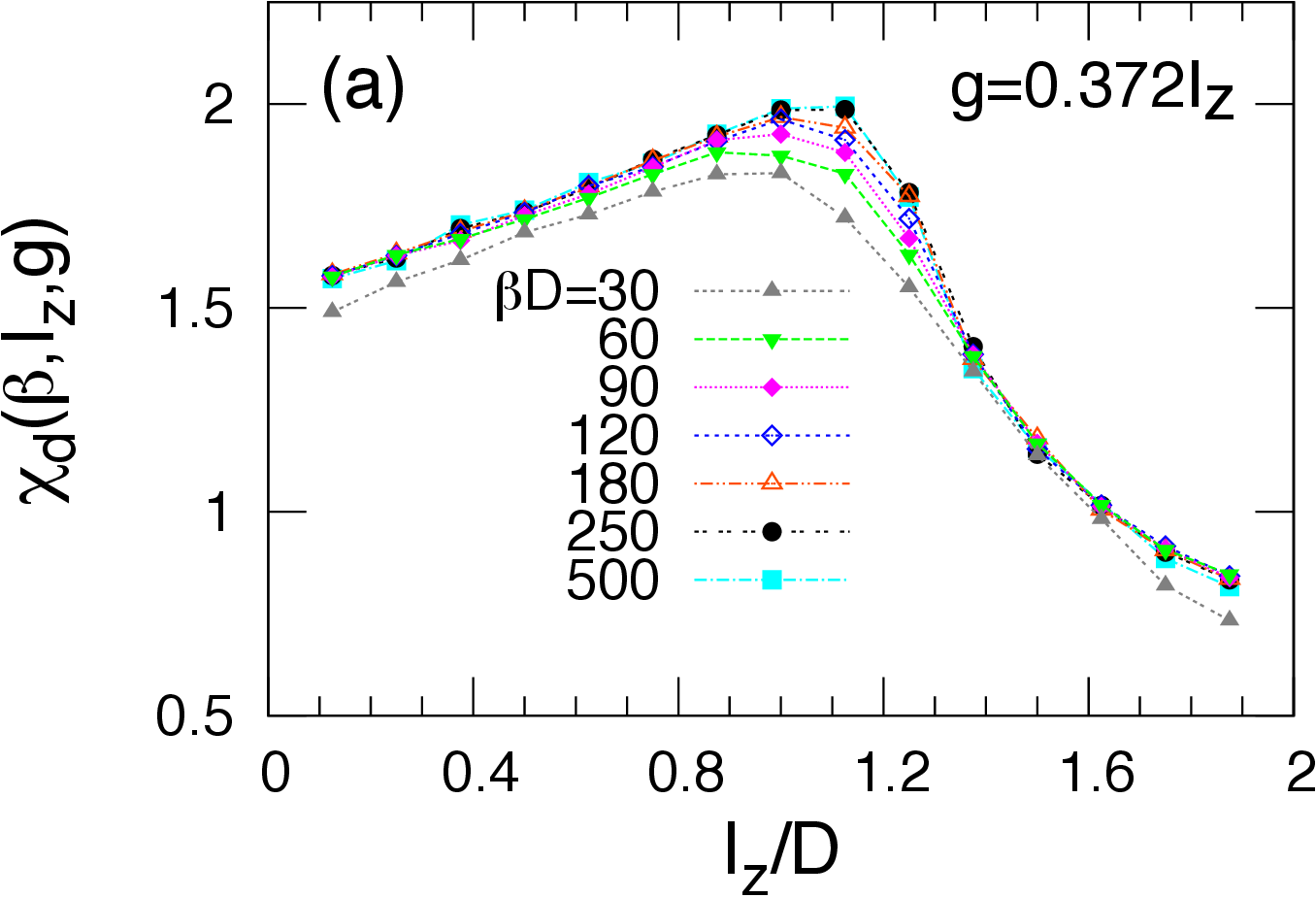}
\end{minipage}
\begin{minipage}[b]{10pc}
\includegraphics[height=1.15in]{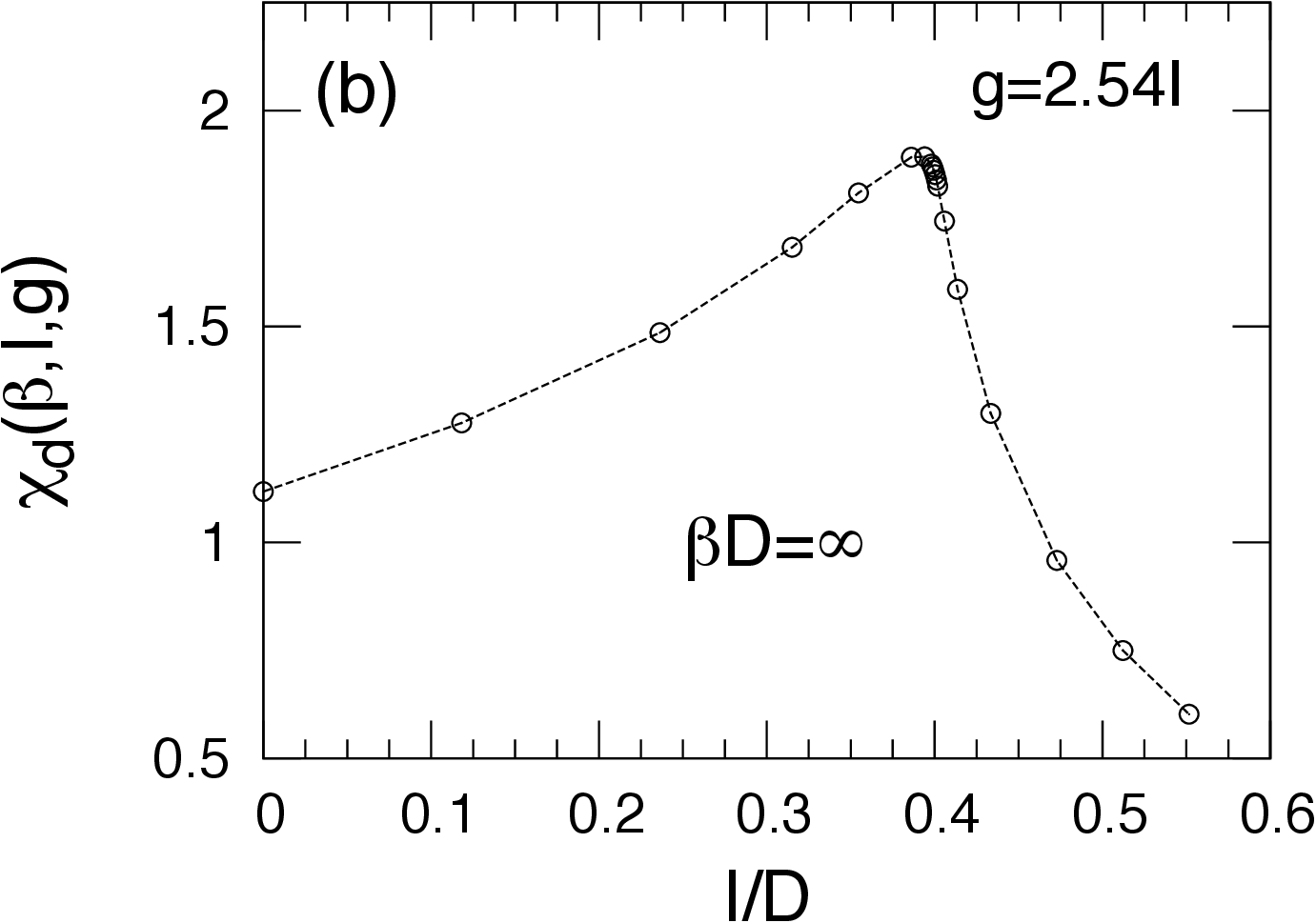}
\end{minipage}
\caption{\label{fig:pairing}(Color online)
(a) Static singlet pairing susceptibility $\chi_d(T,I_z,g)$ vs
$I_z$
for an
Ising $H_{12}$ with $s=0.8$ along the line $g=0.372 I_z$, which crosses the
Kondo-LM phase boundary at $I_z^c=1.25 D$.
(b) Static singlet pairing susceptibility
$\chi_d(T=0,I,g)$
vs $I$
for a Heisenberg $H_{12}$ with $s=0.6$ along the line $g=2.54 I$, which
crosses the Kondo-LM phase boundary at $I_c=0.40 D$. In both (a) and (b),
$\chi_d$ is peaked just on the Kondo side of the phase boundary and
remains elevated at the QCP over its value for $g=I_z=I=0$.}
\end{figure}

More interesting is the singlet susceptibility, which we illustrate along paths
on the $g$-$I_z$ and $g$-$I$ phase diagrams that start from $g=I_z=I=0$ and
cross the Kondo-LM boundary. In C-EDMFT \cite{Pixley.14}, such trajectories are
representative of tuning spin-spin interactions within the lattice model.
Figure \ref{fig:pairing}(a) plots $\chi_d$ vs Ising exchange coupling at a
sequence of temperatures along the cut $g=0.372 I_z$. The pairing susceptibility
grows as $I_z$ increases from zero, is peaked for $I_z$ slightly below $I_z^c$,
and then falls off within the LM phase as the $d$ electrons localize and
decouple from the conduction band. The singlet pairing susceptibility saturates
at temperatures $T\lesssim 0.003 T_K$.

Figure \ref{fig:pairing}(b) illustrates the Heisenberg form of $H_{12}$,
plotting the $T=0$ singlet pairing susceptibility vs $I$ along a
path $g=2.54 I$ that crosses the Kondo-LM boundary. Very much as in the Ising
case, $\chi_d$ rises from $I=0$ and peaks just below $I=I_c$.

The enhancement of the static singlet pairing susceptibility near a
Kondo-destruction QCP is one of the principal results of this work. Although
$\chi_d$ peaks just inside the Kondo phase, the pairing correlation at the
QCP is significantly higher than at $g=I_z=I=0$. We stress that these results
are associated with the critical destruction of the Kondo effect. They differ
from those for $g=0$, where for Heisenberg exchange $\chi_d(T=0)$ diverges at
the Kondo-IS QPT \cite{Zhu}. We have  found (by following the path $g=0.717 I$,
not shown) that the singlet pairing susceptibility also diverges on crossing
the Kondo-IS boundary at some $g>0$, consistent with the picture that this
boundary is governed by the $g=0$ critical point.

The models considered here have both a dynamic (induced by $g$) and a static
($I_z$ or $I$) exchange interaction between the impurities. The combination of
the two antiferromagnetic interactions is responsible for both, the existence
of a Kondo-destruction QCP and the enhancement of $\chi_d$ in its vicinity.
This behavior is likely to have significant effects in lattice systems.
Within C-EDMFT~\cite{Pixley.14}, the cluster pairing susceptibility determines
the lattice pairing susceptibility, in such a way that the enhanced $\chi_d$
may give rise to a pairing instability near a Fermi-surface-collapsing QCP of
a Kondo lattice \cite{Si.01+Si.03,Coleman.01}. As such, this would represent a
new mechanism for superconductivity in the vicinity of antiferromagnetic order,
and would be of considerable interest in connection with the superconductivity
observed in the Ce-115 materials \cite{Hegger00} and related heavy-fermion
superconductors \cite{SiPashchen13}.

In summary, we have introduced and solved two variants of the two-impurity
Bose-Fermi Anderson model using robust numerical methods.
We have mapped out the phase diagrams for these models and shown that each
possesses a line of Kondo-destruction QCPs that are insensitive to
breaking particle-hole symmetry. The QCPs in the two models
belong to the same universality class despite the differing symmetries of the
interimpurity exchange interaction. Just as importantly, we have shown that
the Kondo-destruction quantum criticality in these models \emph{enhances}
singlet pairing correlations. Our results hold promise for elucidating the
superconductivity observed in heavy-fermion metals whose normal state shows
characteristics of Kondo-destruction quantum criticality.

We acknowledge useful discussions with Stefan Kirchner, Lijun Zhu, Aditya
Shashi, and Ang Cai. This work was supported in part by NSF Grants No.\
DMR-1309531 and No.\ DMR-1107814, Robert A.\ Welch Foundation Grant
No.\ C-1411, the East-DeMarco fellowship (JHP), and the
Alexander von Humboldt Foundation. Computer time and IT support at Rice
University was supported in part by the Data Analysis and Visualization
Cyberinfrastructure funded by the NSF under Grant No.\ OCI-0959097.
J.H.P.\ acknowledges the hospitality of the Max Planck Institute for the
Physics of Complex Systems, and Q.S.\ acknowledges the hospitality of the
the Karlsruhe Institute of Technology, the Aspen Center for Physics (NSF Grant
No.\ 1066293), and the Institute of Physics of Chinese Academy of Sciences.


\newpage

\onecolumngrid
\setcounter{figure}{0}
\makeatletter
\renewcommand{\thefigure}{S\@arabic\c@figure}
\setcounter{equation}{0} \makeatletter
\renewcommand \theequation{S\@arabic\c@equation}

\section*{SUPPLEMENTARY MATERIAL -- 
Pairing Correlations Near a Kondo-Destruction Quantum Critical Point}

by: J.\ H.\ Pixley, Lili Deng, Kevin Ingersent, and Qimiao Si

\vskip 1.0 cm

In this supplementary material, we describe in more detail the numerical methods used and present the
renormalization-group (RG)
flow of the two impurity Bose-Fermi Anderson models.

\subsection{Methods}

Two approaches have been used for our study. The first is an extension of the CT-QMC approach \cite{Gull,Pixley.10,Pixley.13}, for the model with an Ising form of the exchange interaction.
After a generalized Lang-Firsov
transformation \cite{Pixley.10,Pixley.13}, the CT-QMC performs time-dependent
perturbation theory in the hybridization and stochastically sums the resulting
series via a Monte-Carlo algorithm. In order to locate the $T=0$ transition via
calculations performed at $T\equiv 1/\beta > 0$, we compute the staggered Binder
cumulant \cite{Binder.81,pixley,Pixley.13} 
$U_4^s(\beta,I_z,g)=\langle M_s^4 \rangle/\langle M_s^2\rangle^2$,
where the staggered magnetization
$M_s = \beta^{-1} \int_0^{\beta} d\tau\,S_s^z(\tau)$ with
$S_s^z=\half(S_1^z-S_2^z)$.
We also calculate the staggered static spin susceptibility
$\chi_{s}(T) = \beta\langle M_s^2 \rangle$.
In the presence of a bosonic coupling to $S_s$, the Heisenberg form of
$H_{12}$ is beyond the reach of the CT-QMC.

The second approach is the Bose-Fermi extension
\cite{Glossop.05} of the NRG \cite{Wilson.75}.
The staggered spin
susceptibility is calculated as
$\chi_s(T)=-\lim_{H_s\to 0}\langle S_s^z\rangle/H_s$ with an additional
Hamiltonian term $H_s S_s^z$. The static pairing correlations are obtained by
Hilbert transformation of the imaginary part of the dynamical susceptibilities,
computed on the real frequency axis in the usual manner \cite{Bulla.08}. The
NRG results presented here were obtained using discretization parameter
$\Lambda=9$, allowing up to 6 bosons per site of the Wilson chain, and keeping
up to 1\,300 many-body eigenstates after each iteration.Ó

\subsection{RG Flow}

\begin{figure}[h]
\includegraphics[height=2in]{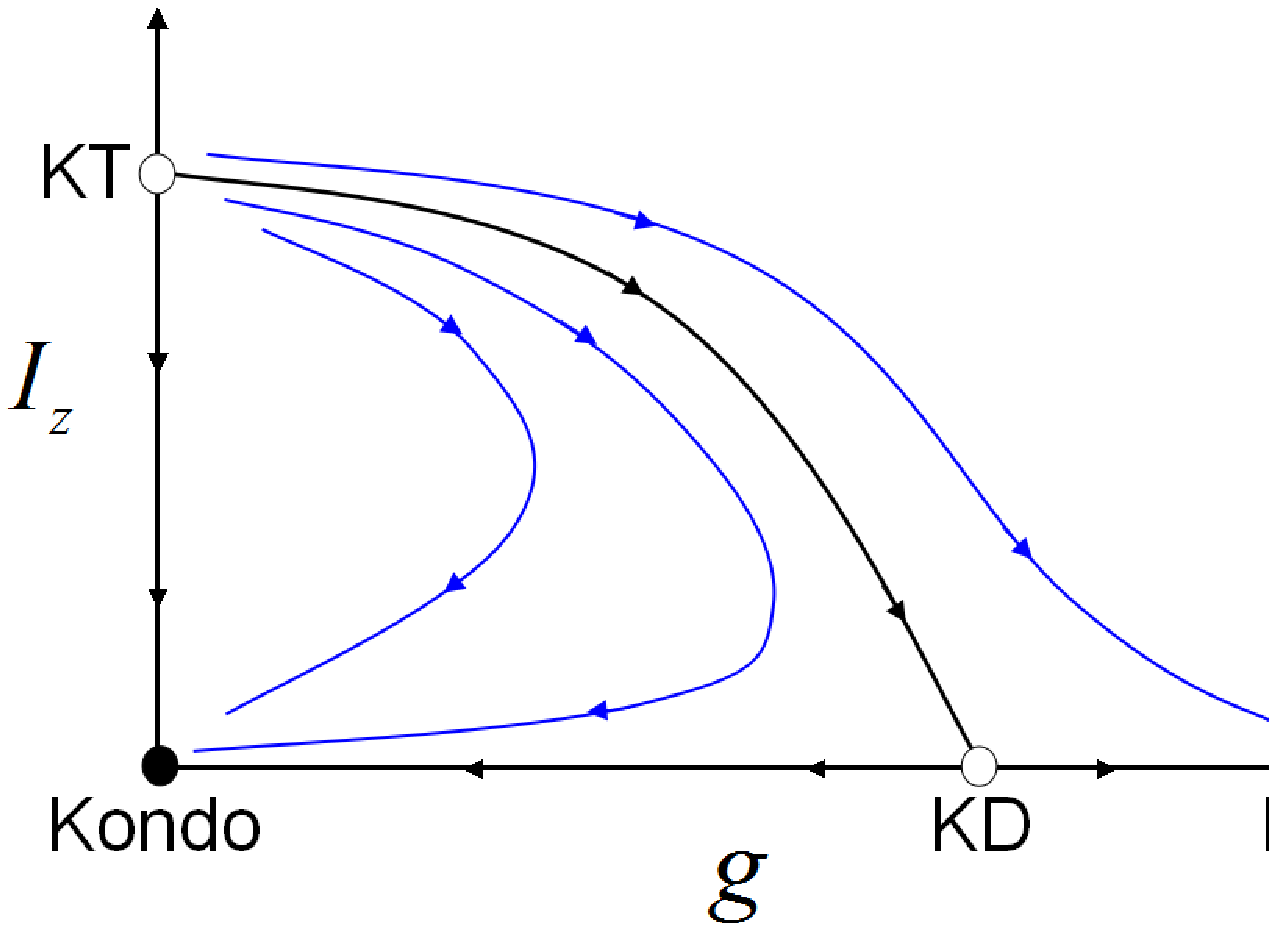}
\caption{(color online).
Schematic
RG
flow on the $g$-$I_z$ plane for the two-impurity Bose-Fermi Anderson model
with Ising
interimpurity
exchange $H_{12}$. Trajectories with arrows represent the flows of the
couplings ($g$ and $I_z$) with the decrease of energy.}
\label{fig:flow_Ising}
\end{figure}

Figure~\ref{fig:flow_Ising} corresponds to the case with Ising
interimpurity
exchange $H_{12}$. 
There are two stable fixed points, indicated by filled circles,
which govern
the Kondo-screened (Kondo)
phase and the local-moment (LM) phase. The Kondo fixed point is at
$(g,I_z)=(0,0)$, while the LM fixed point is at $(g,I_z)=(\infty,0)$. On the
phase boundary, the RG flow is from the Kosterlitz-Thouless (KT) fixed point
toward the Kondo-destruction (KD) fixed point.
The
KT and KD fixed points are both unstable and are shown by open circles. The
direction of the RG flow on the phase boundary reflects
the discussion
in the main text, \textit{i.e.}, all the points for
nonzero $g$ on the phase boundary have the same critical behavior as the KD QCP
on the $g$ axis.

\begin{figure}[h]
\includegraphics[height=2.0in]{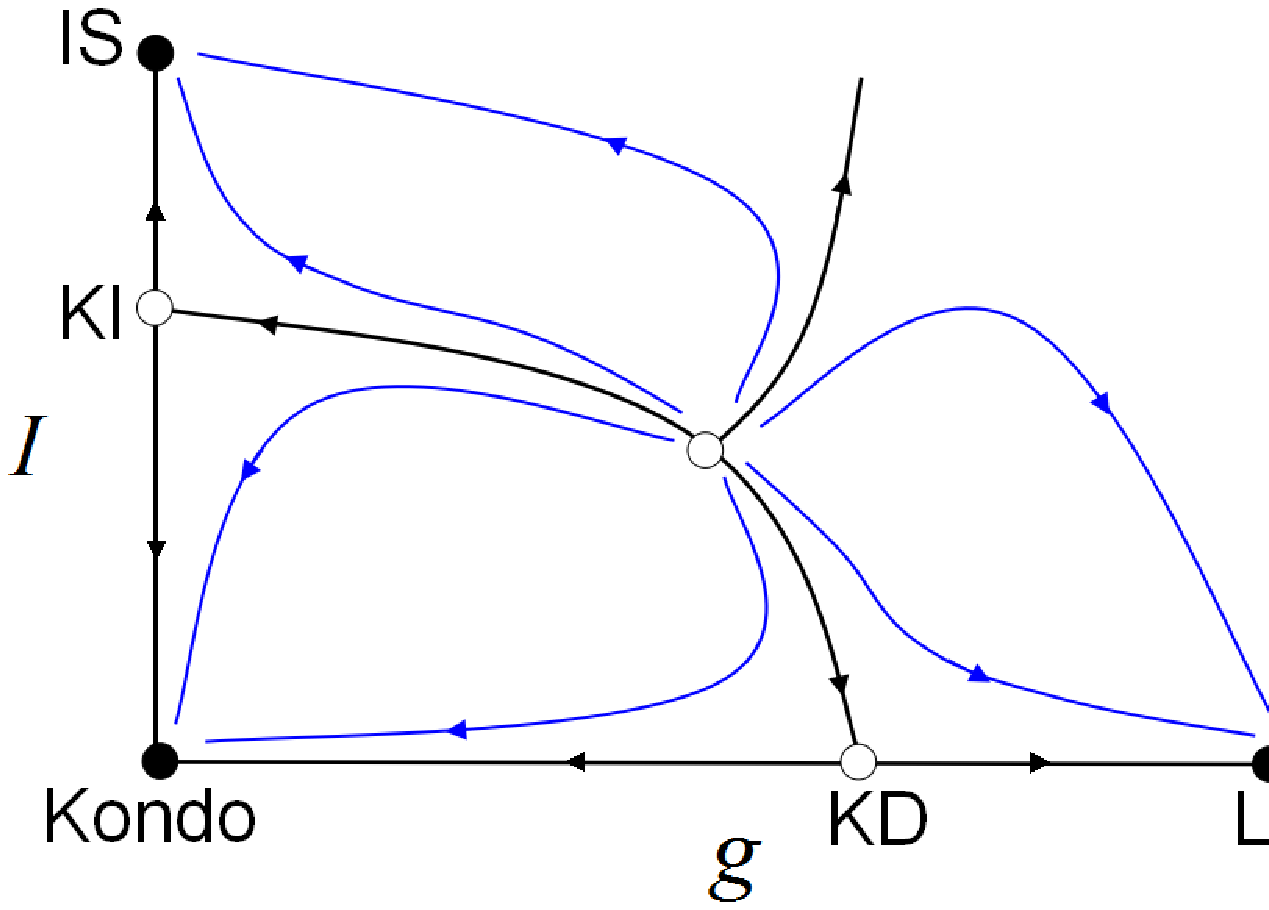}
\caption{(color online).
Schematic RG flow on the $g$-$I$ plane for the two-impurity Bose-Fermi
Anderson model with Heisenberg inter-impurity exchange $H_{12}$.}
\label{fig:flow_su2}
\end{figure}

Figure~\ref{fig:flow_su2} shows the RG flow for the case with Heisenberg
interimpurity
exchange $H_{12}$.
In this case, there are three stable fixed points corresponding to the three
phases. The Kondo-screened (Kondo), local-moment (LM) and interimpurity-singlet
(IS) phases are respectively located at $(g,I)=(0,0)$,  
$(g,I)=(\infty,0)$ and $(g,I)=(0,\infty)$, and are marked by
filled circles. 
On the Kondo-LM phase boundary,
the RG flow is from the triple-point (where the three phases meet) to the
Kondo-destruction (KD) fixed point. Likewise, the RG flow on the Kondo-IS phase
boundary is from the triple-point toward the fixed point KI, which
separates the Kondo and IS phases on the $I$-axis.


\end{document}